\documentclass[pra,a4paper,preprintnumbers,aps,superscriptaddress,twocolumn,floatfix,showpacs,showkeys]{emulateapj}
\usepackage{latexsym}
\usepackage{epsfig}
\usepackage{amssymb}

\newcommand{\bea}{\begin{eqnarray}}
\newcommand{\eea}{\end{eqnarray}}
\newcommand{\be}{\begin{equation}}
\newcommand{\ee}{\end{equation}}

\begin{document}
\title{Accelerating Cosmologies with an Anisotropic Equation of State}
\author{Tomi Koivisto}
\affil{Helsinki Institute of Physics, P.O. Box 64, FIN-00014 Helsinki, Finland}
\author{David F. Mota}
\affil{Institute for Theoretical Physics, University of Heidelberg, 69120 Heidelberg,Germany}
\begin{abstract}
If the dark energy equation of state is anisotropic, the expansion rate of the 
universe becomes direction-dependent at late times. We show that such models are not only cosmologically viable but that 
they could explain some of the observed anomalies in the CMB. The possible anisotropy can then be constrained by studying its effects on the 
luminosity distance-redshift relation inferred from several observations. A vector field action for dark energy is also presented as an example of 
such possibility.
\end{abstract}
\maketitle
\section{Introduction}
It is not clear whether the large-angle anomalies in the observed cosmic microwave 
background are of a cosmological origin and not due to systematics \citep{Eriksen:2003db,Land:2005ad,Copi:2005ff}. 
Nevertheless, it is tempting to associate the apparent statistical anisotropy with dark energy,
since the anomalies occur at the largest scales, and these enter inside the horizon
at the same epoch that the dark energy dominance begins. 

The paramount characteristic of dark energy is its negative pressure. We examine then the possibility that this pressure varies with the direction.
Then also the universal acceleration becomes anisotropic, and one would indeed see otherwise unexpected effects at 
the smallest multipoles of CMB. 
At the background level we will find a similar CMB pattern as in an universe which is ellipsoidal at the era of last 
scattering\citep{Campanelli:2006vb}. We then show that the supernovae could be used to distinguish these different 
scenarios and to constrain the possible anisotropic properties of dark energy. 

There are several motivations for anisotropic models of dark energy. For
instance,  many dark energy models are in principle 
compatible with the FLRW metric but exhibit anisotropic stresses at the perturbative level,
including nonminimally coupled fields, viscous fluids and modified gravity models
\citep{Koivisto:2005mm,uzan,mota,tomi1,tomi2}. It is possible that a more accurate
description of such models should take into account the anisotropic effects on the background expansion, 
which then breaks the statistical isotropy of perturbations too (as appears to have
happened in the CMB \citep{Eriksen:2003db,Land:2005ad,Copi:2005ff}).  
Also, considering dark energy as an effective description of a backreaction of sizable 
inhomogeneities in dark matter \citep{backreaction,back}, the validity of a perfect fluid description could seem 
dubious. 

To have a general description of an anisotropic dark energy component, we consider a phenomenological parameterization of dark energy
in terms of its equation of state ($w$) and two skewness parameters ($\delta$, $\gamma$) and include also a 
coupling term ($Q$) between dark energy and a perfect fluid (dark matter). 
Previous studies anisotropic dark energy have mainly considered the anisotropic properties of the inhomogeneous perturbations 
\citep{Koivisto:2005mm, Battye:2006mb} 
whereas our approach here is to focus on a smooth cosmology with the anisotropic pressure field. Whereas in a
previous approach the anisotropy was only weakly constrained \citep{mota}, we find that in the present 
description only a narrow parameter range can survive the observational 
tests. 
Similar anisotropic inflation has been considered in the early universe \citep{Burd:1991ew,Ford:1989me}. 
Therefore we focus on the present paper more on the dark energy era, where the qualitative differences are that 
matter cannot be neglected and that the cosmologies do not isotropize.
As a proof of a concept, we write also an explicit field theory example where a vector field drives the anisotropic 
acceleration of the universe. 

An anisotropic expansion is not compatible with the 
Robertson-Walker (RW) metric. Hence we use the B(I) (Bianchi type I)  metric which 
generalizes the flat RW metric and may be employed 
to obtain limits on cosmological skew pressures from the CMB\citep{Barrow:1997sy}.
The line-element of a B(I) universe is:   
\be \label{metric}
ds^2 = -dt^2 +a^2(t)dx^2+b^2(t)dy^2 + c^2(t)dz^2.
\ee
There are thus three scale factors, and consequently three expansion rates. In principle all these
could be different, and in the limiting case that all of them are equal one  recovers the
RW case. It is useful to express the mean expansion rate as an average Hubble rate $H$
\be
H \equiv \frac{1}{3}\left(\frac{\dot{a}}{a}+\frac{\dot{b}}{b} + \frac{\dot{c}}{c}\right),
\ee
(where an overdot means derivative wrt $t$) and then the differences of the expansion rates 
as the Hubble-normalized shear $R$ and $S$ \citep{Barrow:1997sy}
\be
R  \equiv \frac{1}{H}\left(\frac{\dot{a}}{a}-\frac{\dot{b}}{b}\right), \quad
S  \equiv \frac{1}{H}\left(\frac{\dot{a}}{a}-\frac{\dot{c}}{c}\right).
\ee
We consider a universe filled with a perfect fluid having the energy-momentum tensor 
$
T^\mu_{(m)\nu} = diag(-1,w_m,w_m,w_m)\rho_m,
$
and dark energy, which we allow to have the most general energy-momentum tensor compatible with 
the metric (\ref{metric})
\be
T^\mu_{\phantom{\mu}\nu} = diag(-1,w,w+3\delta,w + 3\gamma)\rho.
\ee
The generalized Friedmann equation may be written as
\be
H^2 = \frac{8\pi G}{3}\frac{\rho_m+\rho}{1-\frac{1}{9}\left(R^2+ S^2 - RS\right)}.
\ee
We let the two components also interact. The continuity equations are then
\be
\dot{\rho}_m + 3H\left(1+w_m\right)\rho_m = Q H \rho,
\ee
and 
\be
\dot{\rho}   + 3\left[\left(1+w\right)H + \delta \frac{\dot{b}}{b} + 
\gamma\frac{\dot{c}}{c} \right]\rho = -Q H \rho,
\ee
where $Q$ determines the coupling. If it vanishes, it follows that $\rho_m \sim (abc)^{-1-w_m}$ and $\rho \sim 
(abc)^{-1-w}b^{3\delta}c^{3\gamma}$. 
Defining $x \equiv \frac{1}{3}\log{(-g)}$, where the metric determinant $g=-abc$,
one notes that $H = \dot{x}$. We will use $x$ as our time variable rather than $t$. Derivative wrt $x$ is
denoted by star. We also define the dimensionless density fractions
$$
\Omega_m \equiv \frac{8\pi G}{3}\frac{\rho_m}{H^2}, \qquad U \equiv
\frac{\rho}{\rho_m+\rho}.
$$ 
Using $R$, $S$, and $U$ as our dynamical variables, the system can finally be written as
\bea \label{system}
  U^*&=&   U\left(U -1\right)     \left[ \gamma \left(3+R-2S \right)  +\right.\nonumber\\&&\left.
\delta\left(3-2R+S\right) + 3\left(w - w_m\right) \right] - UQ\nonumber,
\\
  S^* & = &
\frac{1}{6}\left(9 - R^2 + RS - S^2 \right) \\ \nonumber
& x & \left\{ S\left[U\left(\delta + \gamma+w-w_m\right)+w_m-1\right]-6\gamma U \right\},
\\ \nonumber
  R^* & = &
\frac{1}{6}\left(9 - R^2 + RS - S^2 \right) \\ \nonumber
& x & \left\{ R\left[U\left(\delta + \gamma+w-w_m\right)+w_m-1\right]-6\delta U \right\}.
\eea
Note the coupling term $Q$ appears only in the evolution 
equation for $U$. Nevertheless its presence can change the dynamics completely \citep{us}. 

\section{Vector Field}

As a proof of concept, we present an explicit field theory for the anisotropically stressed dark energy. 
Consider the vector field action
\be\label{vector}
S = \int d^4 x \sqrt{-g} \left[\frac{R}{16\pi G} -\frac{1}{4}F_{\mu\nu}F^{\mu\nu} - V + qL_m\right]
\ee
where the kinetic term involves $F_{\mu\nu} = \partial_\mu A_\nu - \partial_\nu A_\mu$.
The potential $V$ and the possible couplings $q$ are understood as functions of the quadratic 
term $A^2 = A_\mu A^\mu$. The field equations are
$
G_{\mu\nu} = 8\pi G\left( T^m_{\mu\nu} + T^A_{\mu\nu}\right),
$
where the energy-momentum tensor of the vector field follows by varying wrt the metric
\be \label{emt}
T^A_{\mu\nu} = F_{\mu\alpha}F_{\nu}^\alpha + 2V' A_\mu A_\nu - \left( \frac{1}{4} F_{\alpha\beta} 
F^{\alpha\beta} + V \right) g_{\mu\nu}.
\ee
Prime denotes derivative wrt $A^2$.
With the metric (\ref{metric}), the equation of motion for the $A_0$ component of the field 
dictates that $[q' L_m + V']A_0=0$. Thus we confine to purely spatial vector 
fields. The spatial components of the energy-momentum tensor are,
$$
T^{iA}_{\phantom{i}j} = \frac{1}{a_i^2}\left( -\dot{A}_i\dot{A}_j + 2V' A_iA_j\right) +
        \left(\frac{1}{2}\sum_{k=1}^3 \frac{A_k^2}{a_k^2} - V\right)\delta^i_j.
$$
The off-diagonal terms should vanish. 
Thus one is restricted to consider only vectors
which are parallel to one of the coordinate axis. There, however, could be many such fields.

In the RW case all the diagonal components  of $T^i_{\phantom{i}j}$ should be equal, which
indeed requires three vector fields, one in each coordinate direction, of exactly equal
magnitudes, as the triads of Armendariz-Picon\citep{Armendariz-Picon:2004pm}. With the 
metric (\ref{metric}), the only constraint is that every vector field should be
along a coordinate axis. Any system of these vector fields is thus {\it required} to have 
an anisotropic equation of state, unless it reduces to the very special triad case. Let us therefore briefly
look at the case of a single vector field with $A_1=A_3=0$. 
Notice that, in analogy to system (\ref{system}), the coupling $Q$ is now related to the function $q$. If we 
define the additional dimensionless parameters 
$$
X \equiv \frac{\dot{A}_2}{b H}, \qquad Y \equiv \frac{V}{H^2}, \qquad Y_1 = 2\frac{V'A_2}{b^2 H^2}, 
$$
one can relate the present case to the general notation by noting that
$H^2\rho = X^2/2 + Y$, and that if $S=0$, then  
\be
w = \frac{X^2-2Y}{X^2+2Y}, \quad 
\delta = \frac{-X^2-Y_1}{\frac{3}{2}X^2 + 3Y}. 
\ee
We see that the anisotropy is naturally small if the field is either subdominant or near its minimum.
Such models, even though perturbatively close to standard cosmology, would be excluded by imposing the RW symmetry.
Viable models with large anisotropy do also exist as shown below. An extensive investigation of specific models will 
be undertaken elsewhere \citep{us}.

\section{Scaling solutions (in the axisymmetric case)}

\begin{figure}[ht]
\begin{center}
\includegraphics[width=0.4\textwidth,height=0.4\textwidth]{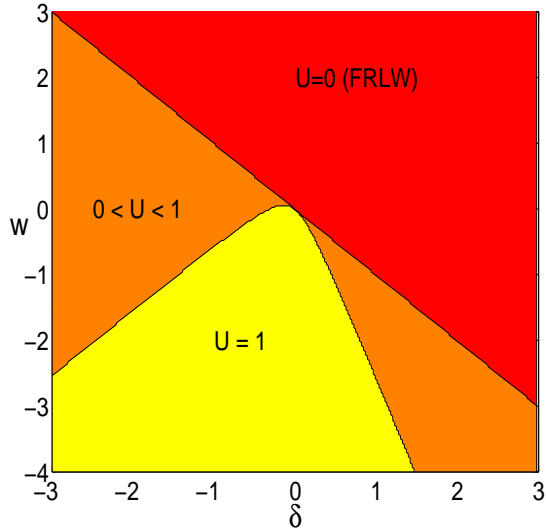}
\caption{\label{outcomes} Asymptotic state  from an Einstein-deSitter stage.
The future fate depends on the dark energy properties $w$ and $\delta$ and is classified into three
possibilities. If $\delta+w > 0$, the isotropically expanding dust 
domination (solution (\ref{flrw_s}) for which $U=0$) 
continues forever. Otherwise, the universe will end up expanding anisotropically 
and either dominated by dark energy (solution (\ref{domi_s}) for which $U=1$) or 
exhibiting a scaling property (solution (\ref{scal_s}) for which $0<U<1$).}
\end{center}
\end{figure}

To begin, we will assume for simplicity that 1) the perfect fluid is 
minimally coupled dark matter, $w_m=Q=0$, and further that 2) the skewness together
with the equation of state of dark energy is constant, $\dot{w}=\dot{\delta}=\dot{\gamma}=0$ (see \citep{us} for an extensive analysis with these assumptions relaxed).  
In our model the universe is then initially (close to) isotropic, but is driven to anisotropic expansion
by the skewness of dark energy.  Under the additional simplifying assumption
of axial symmetry ($S=\gamma=0$), in FIG. \ref{outcomes} we indicate the final possible stages of such universe, corresponding to the fixed 
points of the system (\ref{system}).
The universe will then end up in three possible scenarios: (I) The isotropic RW case corresponding to 
\be \label{flrw_s}
R=0, \quad U = 0.
\ee
(II) An anisotropic dark energy dominated solution with
\be \label{domi_s}
R = \frac{6\delta }{w+\delta-1}, \quad U =1.
\ee 
Or (III) a scaling solution,
\be \label{scal_s}
R = \frac{3\left(w+\delta\right)}{2\delta}, \quad
U = 
\frac{-\left(w+\delta\right)}{3\delta^2-2\delta w - w^2}.
\ee
Given that $w$ is negative, this solution accelerates
$w_{eff} \equiv -\frac{2}{3}\frac{H^*}{H}-1<-\frac{1}{3}$, if
$w<\delta<\frac{w}{3}$ or $-\frac{w}{5}<\delta<-\frac{w}{3}$. 
Note that in the $R=0$ case scaling
solutions could only be found for coupled components.

Notice that within the RW universe it has been proven difficult to address the coincidence problem by finding a model 
entering from a matter-dominated scaling solution to an accelerating scaling solution.
Allowing for the presence of three expansion rates opens up the possibility of  
describing a universe entering from a perfect fluid dominated scaling to 
an anisotropically accelerating scaling era. This might eventually help to understand the 
coincidence problem, since then matter and dark energy would have had similar energy densities
both in the past and in the future (in the past the dark energy fraction, if constant 
should not exceed about $1/10$ \cite{georg}).

We now study the observational implications of models with nonzero 
skewness, relaxing the assumption 2). The relevant case is then a universe entering from RW solution 
(\ref{flrw_s}) to an anisotropically accelerating universe (which generalizes (\ref{domi_s}) 
if $\gamma \neq 0$).

\section{CMB anisotropy}

The B(I) background predicts a quadrupole pattern in the CMB. 
It is not clear how the fluctuations in the photon temperature
distribution evolve when the background expansion becomes anisotropic, 
since the anisotropic sources, even if perturbatively small, couple to the 
perturbations at the first order \citep{us}.
It seems inevitable that statistically anisotropic features would then be, in principle,
present in the whole spectrum of fluctuations (though in our case such features
would be confined to larger scales as the anisotropy of the expansion becomes
important only recently). Several large-angle anomalies have been reported
in the CMB \citep{Eriksen:2003db,Land:2005ad,Copi:2005ff}. It has been demonstrated
that CMB spectrum of the tilted B(VIIh) background \citep{Hawking:1968zw,1985MNRAS.213..917B,Pontzen:2007ii}
would have potential to model these anomalies \citep{Jaffe:2005pw} were it not in conflict with a dark energy 
dominated universe at the smaller angular scales \citep{Jaffe:2005gu}. 

The B(I) model reproduces the predictions of the concordance model at small scales, while 
featuring anomalies at large angles, since the (originally statistically isotropic) CMB field 
experiences an (statistically) anisotropic integrated Sachs-Wolfe effect in the nowadays ellipsoidal 
universe. Furthermore, it is plausible that these anomalies can match the observed ones, since
a B(VIIh) model can be effectively described as a B(I) model with an additional anisotropic 
energy source\footnote{One may obtain the B(VIIO) model from B(I) by adding anisotropic curvature.
To obtain B(VIIh) model one should add also an isotropic curvature similar to such present
in B(V) models.} 

As a first step, we  check whether the background (\ref{metric}) is 
compatible with the available cosmological data. There are a number of calculations of  
CMB anisotropies in general Bianchi universes already available in the
literature \citep{Hawking:1968zw,1985MNRAS.213..917B,Pontzen:2007ii}. Here we consider the dark energy created quadropole and the 
effects to the luminosity distance.

By considering the geodesic equation for photons, one can derive an equation for the redshift $z$ of a 
photon arriving from the a direction $\hat{{\bf p}}$ 
\be 
\label{z_ellips}
1+z(\hat{{\bf p}}) = 
\frac{1}{a}\sqrt{1 + \hat{p}_y^2e_y^2 + \hat{p}_z^2e_z^2}
\ee
in terms of the eccentricities
\be \label{ellips}
e_y^2 = \left(\frac{a}{b}\right)^2-1, \quad
e_z^2 = \left(\frac{a}{c}\right)^2-1.
\ee
Note that the scale factors and eccentrities here are evaluated at the time of 
last scattering in the case that the scale factors are all normalized to unity
today. If $T_*$ is the temperature at decoupling,
the temperature field is given by $T(\hat{{\bf p}}) = T_*/(1+z(\hat{{\bf p}}))$,
and its spatial average is $4\pi\bar{T} = \int d\Omega_{\hat{{\bf p}}}T(\hat{{\bf p}})$.
The anisotropy field is then
\be
\frac{\delta T(\hat{{\bf p}})}{\hat{T}} = 1- \frac{T(\hat{{\bf p}})}{\bar{T}}. 
\ee
The coefficients in the spherical expansion of this anisotropy field are called
$a_{\ell m}$, and due to orthogonality of spherical harmonics $Y_{\ell m}$ are given
by
\be
a_{\ell m} = \int d\Omega_{\bf p} \frac{\delta T(\hat{{\bf p}})}{\hat{T}} Y^*_{\ell m}.
\ee
The multipole spectrum is described by 
\be
Q_\ell = \sqrt{\frac{1}{2\pi}\frac{\ell(\ell+1)}{(2\ell+1)}\sum_{m=-\ell}^{\ell}|a_{\ell m}|^2}.
\ee
Expanding the redshifts (\ref{z_ellips}) in powers of the eccentricities (\ref{ellips}) one 
notes that the $a_{\ell m}$ will be real (since there is only even dependence on 
the polar angle in the anisotropy field and the imaginary parts of the $e^{i m\phi}$ integrate 
to zero), and that for all odd $\ell$ the $a_{\ell m}$ will vanish (since only even powers
of the azimuthal angle appear in the expansion). To first order in $e_{y,z}^2$, 
$\bar{T} = aT_*[1-\frac{1}{6}(e_z^2+e_y^2)]$, and in addition to the monopole, there is only 
the quadrupole
\bea \label{quad}
a_{20}&=&\frac{1}{3}\sqrt{\frac{\pi}{5}}\left(2e_z^2-e_y^2\right), \qquad
a_{21}=a_{2-1}= 0, \\ a_{22}&=&a_{2-2}= -\sqrt{\frac{\pi}{30}}e_y^2
\quad \Rightarrow \quad 
Q_2 = \frac{2}{5\sqrt{3}}\sqrt{e_z^4+e_y^4-e_z^2e_y^2}
\nonumber
\eea

\begin{figure}[ht]
\begin{center}
\includegraphics[width=0.4\textwidth,height=0.4\textwidth]{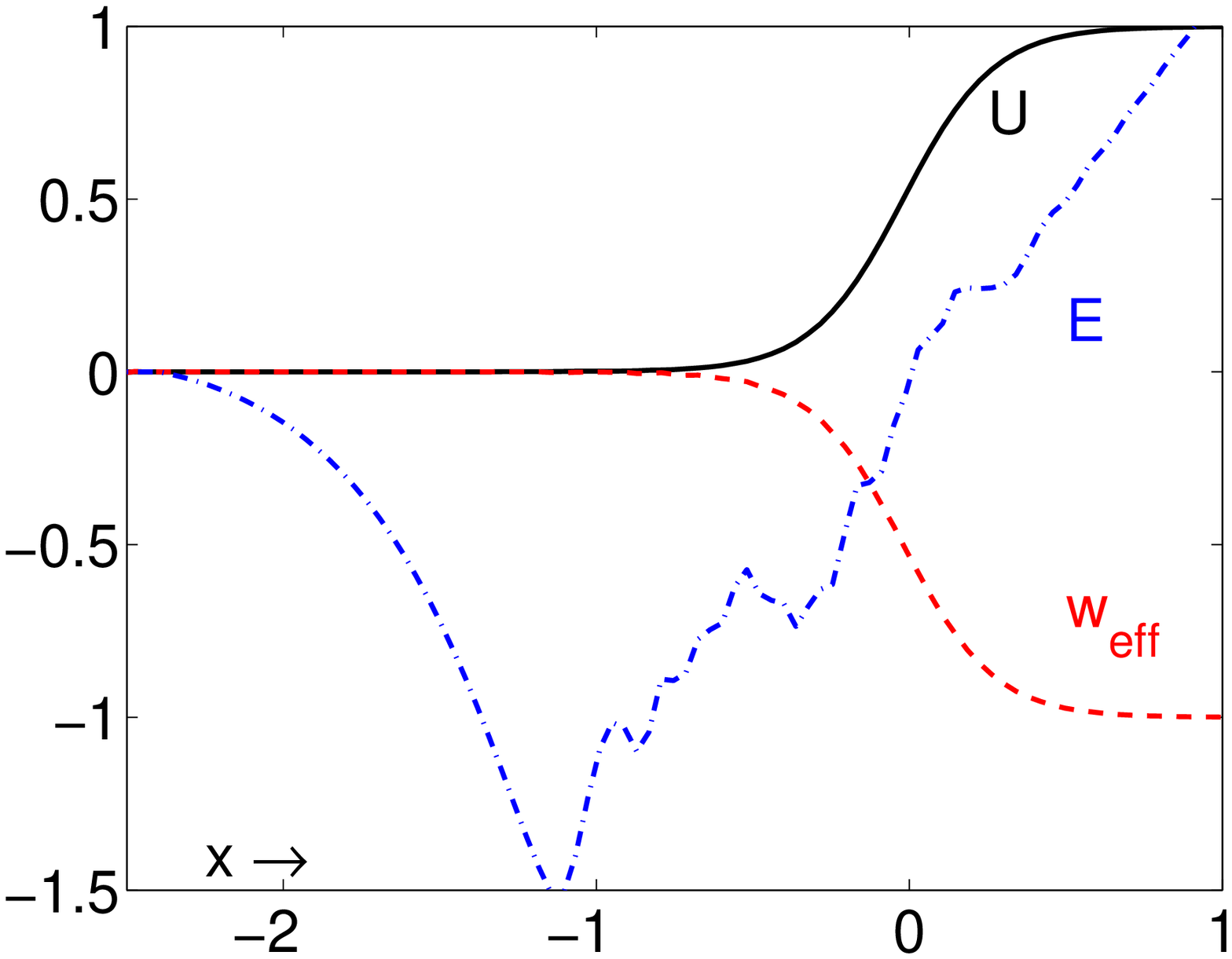}
\caption{\label{tuningpic} A minimally coupled vector field satisfying the quadrupole
constraint. The solid (black) line is the dark energy density fraction $U$, the dashed (red) line is the 
effective equation of state of the universe $w_{eff}$, and the dash-dotted (blue) line describes the 
evolution of eccentricity, $E=500e_y^2$. The potential is a double-power law $V = m A^2 + \lambda A^{-4}$. The 
dynamics of the field is such that though there are significant anisotropies, the eccentricity at the 
present is close to zero. This may be achieved with different power-law potentials, but requires fine-tuning 
of the mass scales $m$ and $\lambda$.}
\end{center}
\end{figure}

The observed value of $Q_2$ is lower than the concordance model predicts \citep{Hinshaw:2006ia}. It has been suggested 
in previous works  that this discrepancy could be explained by an ellipsoidality of 
the universe \citep{Campanelli:2006vb,BeltranJimenez:2007ai}. This would require that the 
anisotropy of the background is suitably oriented with respect to the intrinsic quadrupole and cancels its power to 
a sufficient amount. Too large anisotropy would of course only make the situation worse regardless of the orientation.
Depending on the cosmological model, one should have $Q_2 \lesssim 2\cdot 10^{-5}$ to be consistent with observations 
taking into account the cosmic variance. The constraints this implies on the skewness of dark energy are very tight. 
However, we remark that in more general models, in particular with time-varying $\delta$ and $\gamma$, one could 
allow more
anisotropy. It is in principle possible for arbitrarily anisotropic expansion to escape 
detection from CMB (considering only the effects from the background), as long as the
expansion rates evolve in such a way that $e_z=e_y=0$. In other words, the (background) 
quadrupole vanishes, if each scale factor has expanded - no matter how anisotropically -
the same amount since the last scattering. An example of such a scenario, derived from the action (\ref{vector}), 
is shown in FIG. \ref{tuningpic}.

\section{SNIa luminosities}

\begin{figure}[ht]
\begin{center}
\includegraphics[width=0.4\textwidth,height=0.4\textwidth]{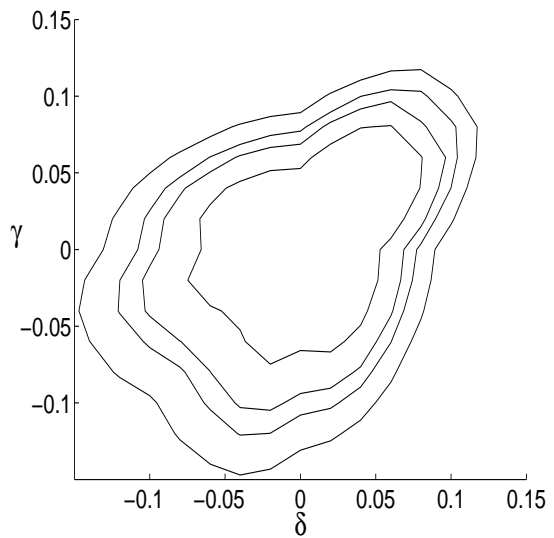}
\caption{\label{eoslims} Limits on the (constant) skewness parameters 
of dark energy arising from the SNIa data, when when $\Omega_m=0.3$ and $w=-1$. The contours correspond to
68.3, 90, 95.4 and 99 percent confidence limits. }
\end{center}
\end{figure}

\begin{figure}[ht]
\begin{center}
\includegraphics[width=0.4\textwidth,height=0.4\textwidth]{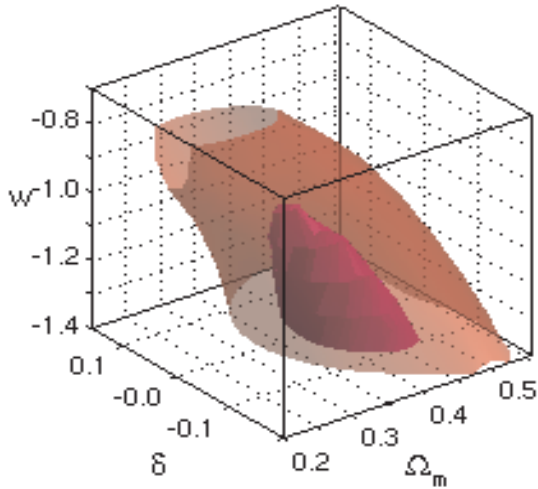}
\caption{\label{eoslims2} 
Constraints arising from the SNIa in the axisymmetric case $\gamma=0$. Inside the darker
isosurface, the fit is as good as in the $\Lambda$CDM model, $\chi^2 < 158$. Inside the lighter
isosurface, one has $\Delta\chi^2 < 8.02$. One notes that larger skewness $\delta$ would typically be
compatible with the SNIa data for phantom equations of state $w<-1$ and large matter densities.}
\end{center}
\end{figure}

The luminosity-redshift relationship of the SNIa data could be used to probe the possible anisotropies
in the expansion history. This is a useful complementary probe since these objects are observed
at the $z < 2$ region, whereas CMB comes from much further away at $z \sim 1100$. The luminosity distance 
at the redshift $z$ in the direction $\hat{p}$ is now given by \citep{us}
\be \label{lumi}
d_L(z,{\bf \hat{p}}) = (1+z) 
\int_{t_0}^{t(z)}\frac{dt}{\sqrt{\hat{p}_x^2a^2+\hat{p}_y^2b^2+\hat{p}_z^2c^2}}. 
\ee
To test this prediction with the data, we apply the formula (\ref{z_ellips}) for each
observed redshift of a supernova and match its luminosity distance inferred from the observation
to the one computed from (\ref{lumi}). In addition, we have to take into account also the angular 
coordinates of each individual supernovae in the sky to fix ${\bf \hat{p}}$ for each object.
In our analysis we use the GOLD data set \citep{Riess:2006fw}, which consists of five
subsets of data\footnote{
SNIa
Angular coordinates 
of each of the 182 GOLD supernovae
can be found in \citep{Riess:2006fw,Astier:2005qq} and 
http://cfa-www.harvard.edu/ps/lists/Supernovae.html}. We marginalize over the directions
in the sky and over the present value of the Hubble constant. 
The results are summarized in FIGs. \ref{eoslims} and \ref{eoslims2}. The best-fit anisotropic models 
are only slightly preferred over the $\Lambda$CDM, the difference being $\Delta \chi^2 \approx 1$. 
Because of the additional parameters in the anistropic models, the reasonable interpretation of these
statistics is that the SNIa data favors isotropic expansion.     

The SNIa data constrains the skewness parameters much looser than the CMB quadrupole, if they are constant.
However, in specific models with time-evolving $\delta$ and $\gamma$, the constraints from CMB and from
SNIa can be of comparable magnitude and allow anisotropy to an interesting degree.
This means that the even if the CMB formed isotropically at early time,
it could be distorted by the acceleration of the later universe in such a way 
that it appears to us anomalous at the largest scales\footnote{Generating these effects at low redshift has an advantage that  
it relaxes constraints which would otherwise come from the CMB polarization \citep{Pontzen:2007ii} and could be strong for a given 
temperature anisotropy in isotropizing models because of the significant polarization anisotropy at last scxattering.
However, anisotropic dark energy could evade this since the optical depth to $z \sim 1$ is very small.}. The future SNIa 
data, with considerably improved error bars
(e.g. from the SNAP experiment,  http://snap.lbl.gov/.)
on $d_L(z,{\bf \hat{p}})$ might be used to
rule out this possibility, and to distinguish whether the possible statistical anisotropy was 
already there at last scattering or whether it is due to dark energy.

\section{Conclusions}

In conclusion, dark energy with an anisotropic equation of state 
might be the culprit for both 
the cosmic acceleration and the large-angle anomalies in the CMB. This  
might also be the key to understand the coincidence problem. The present SNIa data allows anisotropic acceleration, 
but SNAP could set things straight about the skewness of dark energy and so of its nature. 
Such possibility would  open a completely new window not only on the nature of the CMB anomalies but 
also into high energy physics models beyond the usual isotropic candidates 
of dark energy such as scalar fields or the cosmological constant. 

\acknowledgments
We thank Y. Gaspar, H. Sigbjorn and the anonymous referee for useful comments.
TK acknowledges support from the Magnus Ehrnrooth Foundation, the Finnish 
Cultural Foundation and EU FP6 Marie Curie Research and Training 
Network "UniverseNet" (MRTN-CT-2006-035863). DFM acknowledges support from the A. Humboldt
Foundation and the Research Council of Norway through project number 159637/V30.

\end{document}